\documentclass[letterpaper]{article} 
\usepackage{aaai24}  
\usepackage{times}  
\usepackage{helvet}  
\usepackage{courier}  
\usepackage[hyphens]{url}  
\usepackage{graphicx} 
\def\UrlFont{\rm}  
\usepackage{natbib}  
\usepackage{caption} 
\usepackage{algorithm}
\usepackage{algorithmic}
\usepackage{multirow}
\usepackage{multicol}
\usepackage{amsmath}
\usepackage{newfloat}
\usepackage{listings}
\DeclareCaptionStyle{ruled}{labelfont=normalfont,labelsep=colon,strut=off} 
\floatstyle{ruled}
\newfloat{listing}{tb}{lst}{}
\floatname{listing}{Listing}
\title{2022 Flood Impact in Pakistan: Remote Sensing Assessment of Agricultural and Urban Damage}
\author{
    Aqs Younas\textsuperscript{\rm 1}\equalcontrib,
    Arbaz Khan\textsuperscript{\rm 2}\equalcontrib,
    Hafiz Muhammad Abubakar\textsuperscript{\rm 3},
    Zia Tahseen \textsuperscript{\rm 4},
    Aqeel Arshad \textsuperscript{\rm 4},
    Murtaza Taj\textsuperscript{\rm 2}\equalcontrib,
    Usman Nazir\textsuperscript{\rm 3}\equalcontrib
}
\affiliations{
    \textsuperscript{\rm 1} 46 Labs, Pakistan\\
    \textsuperscript{\rm 2} Syed Babar Ali School of Science and Engineering (SBASSE), Lahore University of Management Sciences, Pakistan \\
    \textsuperscript{\rm 3} Center for AI Research (CAIR), School of Computer and IT, Beaconhouse National University, Pakistan \\
    \textsuperscript{\rm 4} 4XPillars, Germany \\
     aqs.younas@46labs.com, \{23100243, murtaza.taj\}@lums.edu.pk, \{f2021-641, usman.nazir\}@bnu.edu.pk,\\ \{z.tahseen, a.arshad\}@4xpillars.com 
    
}

\begin{document}

\maketitle

\begin{abstract}
   Pakistan was hit by the world's deadliest flood in June 2022, causing agriculture and infrastructure damage across the country. Remote sensing technology offers a cost-effective and efficient method for flood impact assessment. This study is aimed to assess the impact of flooding on crops and built-up areas. Landsat 9 imagery, European Space Agency-Land Use/Land Cover (ESA-LULC) and Soil Moisture Active Passive (SMAP) data are used to identify and quantify the extent of flood-affected areas, crop damage, and built-up area destruction. The findings indicate that Sindh, a province in Pakistan, suffered the most. This impact destroyed most Kharif season crops, typically cultivated from March to November. Using the SMAP satellite data, it is assessed that the high amount of soil moisture after flood also caused a significant delay in the cultivation of Rabi crops. The findings of this study provide valuable information for decision-makers and stakeholders involved in flood risk management and disaster response.
\end{abstract}

%

\section{Introduction}
Flooding has sporadically impacted Pakistan since independence, mostly during the monsoon season from July to September. Between 1950 and 2011, Pakistan experienced a total of 21 significant floods, occurring at an approximate rate of one flood every three years. These calamities resulted in the tragic loss of 8,887 lives, extensive damage or complete destruction of 109,822 villages, and economic losses totaling $\$19$ billion~\cite{ali2013indus}. In 2022, the catastrophic flooding caused approximately $\$14.9$ billion damage, and economic losses of $\$15.2$ billion \cite{worldBank}. Cities of Sindh, Balochistan, and the Southern part of Punjab were mainly affected. In this technological era, the volume and extent of damage caused by floods may be evaluated using Remote Sensing (RS) and Geographical Information Systems (GIS).

Remote sensing is considered the fastest source of acquiring data for flood management(\cite{HaqEtAl}, \cite{UlloaEtAl}, \cite{Zhang}). It is used to assess, identify, and evaluate pre-flood and post-flood phases. With the continuous advancement in remote sensing and artificial intelligence, the effects of floods are now observable through real-time satellite imagery and the utilization of Neural Network Classifiers.

The impact of flood-induced crop and urban damage encompasses food scarcity, economic setbacks, and disruptions in daily life, emphasizing the necessity of proactive flood management and mitigation strategies to address these challenges effectively. This research paper focuses on the impact of the 2022 flood in the Sindh region of Pakistan on crop cultivation, production, and built-up destruction. We will leverage spectral indices computed from Landsat 9 data to transform raw spectral information into meaningful indicators. Subsequently, we plan to employ a Random Forest classifier for land cover classification and analysis. As flooding rates continue to rise, it affects soil moisture levels. Additionally, an analysis is conducted on soil moisture levels between June 2022 and January 2023 to determine post-flood water requirements for Rabi crops.

\begin{figure}[!h]
    \centering
    \includegraphics[width=0.5\textwidth]{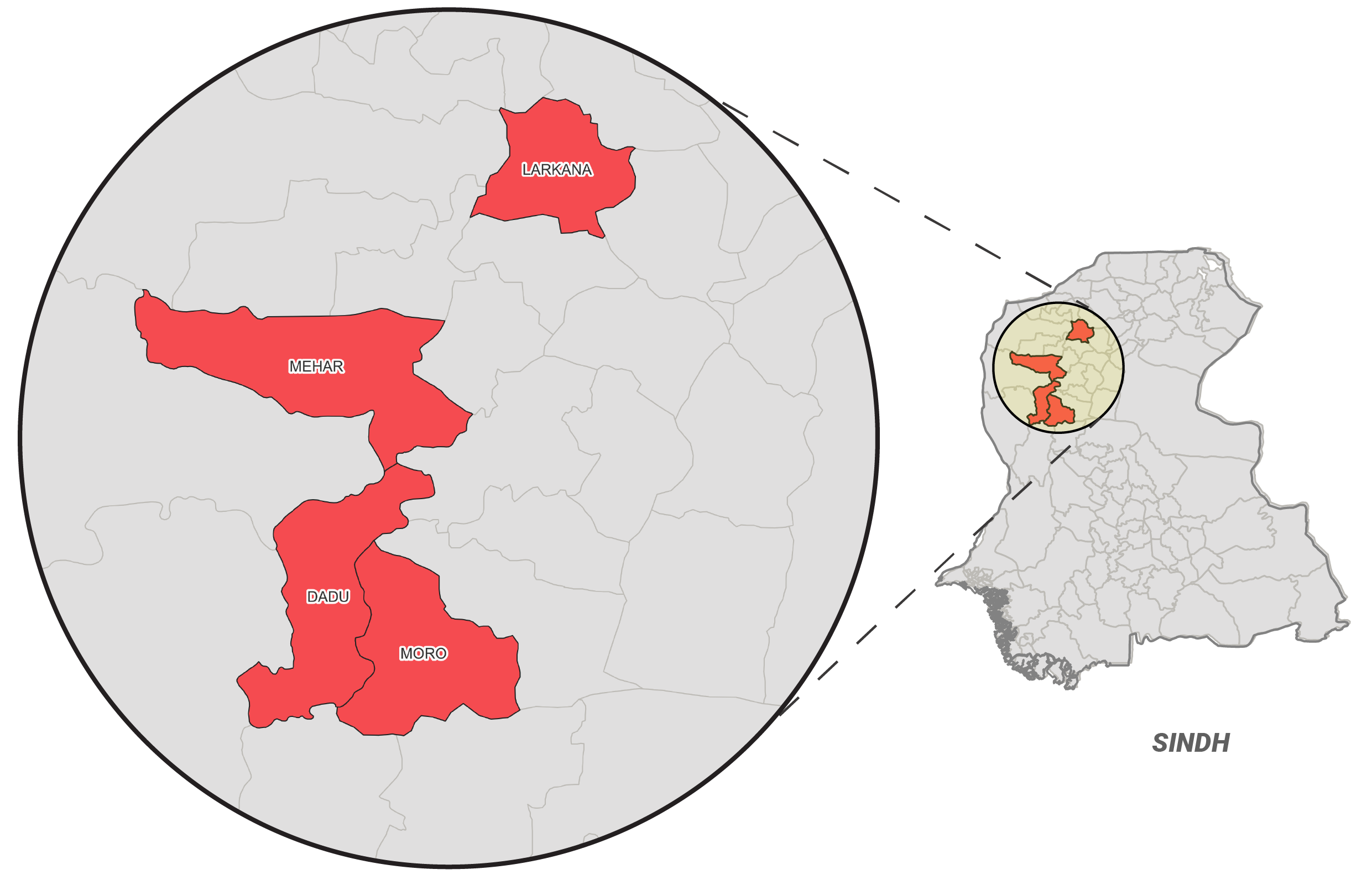}
    \caption{The study area in the visual diagram includes the four flood-affected districts of Mehar, Dadu, Larkana, and Moro in Sindh Pakistan, covering a total of $3,054$ square kilometers.}
    \label{fig:studyarea}
\end{figure}

\begin{figure*}[!h]
    \centering
        \begin{tabular}{c}
        \includegraphics[width=0.75\textwidth]
            {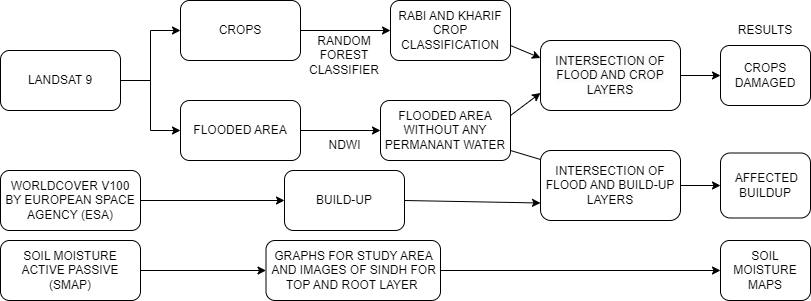}
        \end{tabular}
        \\
    \caption{Visual representation of the methodology.}
    \label{fig:methodology}
\end{figure*}
\subsection{Contribution} This paper makes a significant contribution by conducting a comprehensive assessment of flood impacts on both crops and built-up areas, utilizing two distinct technologies. We employ Landsat-9 in combination with a Random Forest classifier for crop classification, while the ESA-LULC product is utilized for analyzing built-up areas. The precise classification of flooded areas is achieved through the application of the Normalized Difference Water Index (NDWI). By overlaying these layers, we offer a holistic perspective on the overall flood impact. Additionally, we harness SMAP product data to predict floods and evaluate their real-time effects on crop cultivation. The insights derived from this study hold the potential to enlighten policymakers and stakeholders about the critical necessity of enhancing flood management strategies to mitigate the detrimental consequences of such natural disasters on crop yields and overall food security.
\subsection{Climate Impact Assessment}
This paper showcases the detrimental effects of floods on crop cultivation and production in the Sindh region of Pakistan. By utilizing Landsat 9 imagery and soil moisture maps, it provides valuable insights into the extent of damage caused by the floods and their lingering impact on crop production. These findings underscore the importance of developing and implementing effective flood management strategies to safeguard crop yields and ensure food security. 

\section{Literature Review}
Haq et al. (2012) utilized remote sensing and GIS methods, alongside MODIS Terra and Aqua data at a 250m resolution, for flood monitoring and damage assessment in Pakistan's Sindh province \cite{HaqEtAl}, showcasing the value of these technologies in disaster management. Ulloe et al. (2021) demonstrated the improved efficiency of flood mapping using satellite data by achieving a remarkable accuracy of 0.92 through the combined utilization of Synthetic Aperture Radar (SAR) imagery and spatiotemporal simulation framework.

Numerous research endeavors have explored crop classification using spatial imagery and neural network classifiers, as evidenced by the work of Guo et al. (2022) and Qadeer et al. These studies have often relied on crop phenological parameters like the Normalized Difference Vegetation Index (NDVI) and Enhanced Vegetation Index (EVI), which may not always be readily available or precise for every region. Guo et al. (2022) addressed this challenge by introducing a method that utilizes a blend of SAR and optical images for time series analysis \cite{GuoEtAl}. In a complementary approach, Qadeer et al. proposed combining a 3-dimensional Convolutional Neural Network (CNN) and 1-dimensional CNN for large-scale crop classification \cite{Qadeer}.

Recently, the progress in machine learning methodologies has provided an avenue for modeling cause-and-effect relationships such as floods and crops \cite{Lazin}. A number of studies have been conducted to assess the effects of floods on crops using remote sensing and crop condition profiles. Yu et al. (2013) calculated post-flood crop condition indices using high spatial resolution Landsat and MODIS images and crop phenology information. The flood damage was then extracted from crop condition indices, using regression analysis \cite{GenongYu2013}. Lazin et al. (2021) developed a deep learning model based on CNN architecture to predict the area of damage caused by the flooding of croplands. The model was effective in object identification however, it excluded the topographic information and was only feasible for extreme damage events \cite{Lazin}. \\

\begin{figure*}[!h]
    \centering
        \begin{tabular}{c}
        \centering\includegraphics[width=0.7\textwidth]
        {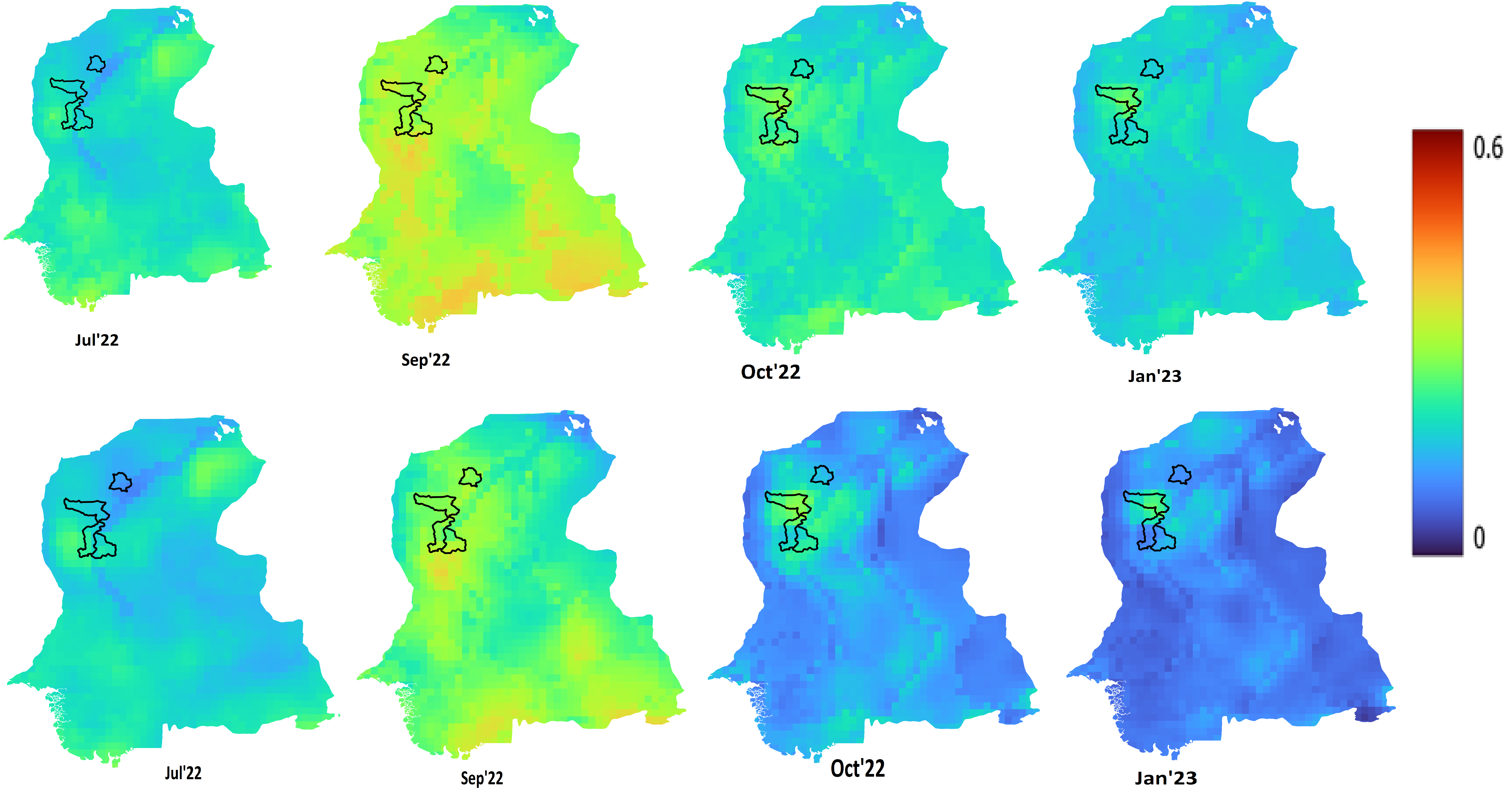}
        \end{tabular}
    \caption{The top row corresponds to the Soil Moisture Index for the root layer, while the second row represents the Soil Moisture Index for the top layer.}
    \label{fig:soil_moisture}
\end{figure*}
\section{Tools and Data Sources}
\begin{itemize}
    \item Landsat 9 Collection 2 Tier 1 calibrated top-of-atmosphere (TOA) reflectance dataset is utilized for flood mapping.
    \item Spectral indices, namely NDWI, NDVI, and EVI, are used for crop classification.
    \item A World cover map at a 10-meter resolution based on Sentinel-1 and Sentinel-2 data by the European Space Agency (ESA) is used for built-up classification.
    \item NASA SMAP level 4 product is used to collect soil moisture maps.
\end{itemize}

\section{Proposed Methodology}
This research paper uses Landsat 9 satellite imagery and Worldcover v100 by the European Space Agency (ESA) to estimate the damage caused by floods in Pakistan in 2022, particularly in the Sindh region, focusing on crop cultivation and production. A Random Forest classifier is applied for crop classification, and then the area of the damaged crops is determined through the intersection of the crop and flood layer. A similar approach is applied to assess the effect of floods on built-up areas. Figure~\ref{fig:methodology} provides a visual representation summarizing the research methodology.

Crop condition profiles are derived from different vegetation indices, including the Normalized Difference Vegetation Index (NDVI) and Enhanced Vegetation Index (EVI) \cite{GuoEtAl}. Flood pixels are derived using the water index, Normalized Difference Water Index (NDWI) \cite{Qadeer}.  
\begin{equation}
f(Lat, Lon) =
\begin{cases}
    1 & \text{if } \text{NDWI} < 0.15 \text{ and } \text{NDVI} < 0.2 \\
    0 & \text{otherwise}
\end{cases}
\end{equation}

where f(.) shows classification decision and (Lat,\ Lon) shows geographical coordinates of specific locations on Earth's Surface. The threshold of $NDWI \le 0.15$ indicates the spectral characteristics of that area are more consistent with non-water features. 

In the initial phase of our study, Landsat-9, an optical satellite with a 16-day imaging cycle, was employed for flood detection and crop classification. We applied the NDWI (Normalized Difference Water Index) to Landsat-9 images taken from April to August. The NDWI threshold helped us delineate flooded areas, excluding permanent water bodies from our analysis.

For crop classification, we utilized Landsat-9 and the Random Forest machine learning algorithm. We focused on Kharif crops in the Lower Bari Doab Canal (LBDC) region of Punjab, using NDVI values and data from the Punjab Irrigation Department (PID)~\cite{irrigation_punjab}. Pre-flood Landsat-9 optical imagery underwent masking to remove areas with low NDVI values, mainly targeting shrubs and grasslands. We then applied a Random Forest classifier to categorize Kharif crops in Sindh's tehsils, providing insights into the agricultural landscape.

\begin{table}[h]
    \centering
        \caption{Built-up and crop area (square kilometers) affected by 2022 floods in Pakistan.}
    \resizebox{0.5\textwidth}{!}{
        \begin{tabular}{ |c|c|c|c| }
        \hline
         \textbf{Tehsil} & \textbf{Total Area} & \textbf{Area of Affected Kharif Crop} & \textbf{Affected Built-Up} \\ 
         \hline
         Mehar & 1020.4 & 129.9 & 0.70\\
         Dadu & 784.5 & 13.8 & 0.10\\  
         Larkana & 499.3 & 27.6 & 0.51\\
         Moro & 750.2 & 19.0 & 0.13\\
         \hline
        \end{tabular}
    }

    \label{fig:crop_damage}
\end{table}

\begin{table*}[h]
    \centering
      \caption{Variation in crop area (square kilometers) production in Pakistan between 2021 and 2022.}
        \begin{tabular}{ |c|c|c|c|c|c|c|c|c|c| }
        \hline
        \multicolumn{2}{|c}{}  & \multicolumn{4}{|c}{2021} & \multicolumn{4}{|c|}{2022} \\
        \hline
         Month & Tehsil & Rice & Maize & Cotton & Orchards & Rice & Maize & Cotton & Orchards \\ 
         \hline
         
         \multirow{4}{4em}{May} & Mehar & 7.10 & 2.63 & 15.47 & 3.258 & 0.35 & 3.79 & 26.21 & 0.03 \\
          & Dadu & 22.36 & 2.90 & 21.82 & 24.72 & 0.57 & 1.97 & 43.00 & 0 \\
          & Larkana & 20.03 & 5.72 & 27.41 & 52.95 & 17.23 & 13.80 & 51.73 & 3.00 \\
          & Moro & 86.70 & 8.27 & 65.30& 134.78 & 8.77 & 32.65 & 229.07 & 0.04 \\
         \hline
         \multirow{4}{4em}{June} & Mehar & 10.23 & 8.11 & 6.59 & 1.01 & 13.71 & 1.91 & 11.99 & 0.185 \\
          & Dadu & 51.62 & 31.37 & 29.02 & 4.55 & 43.12 & 1.67 & 27.12 & 0.72 \\
          & Larkana & 48.38 & 14.81 & 14.56 & 3.58 & 52.00 & 5.69 & 16.34 & 1.29 \\
          & Moro & 103.38 & 109.74 & 125.54 & 36.55 & 177.30 & 11.66 & 96.40 & 9.68 \\
          \hline
         \multirow{4}{4em}{Jul-Sep} & Mehar & 465.88 & 53.18 & 47.04 & 7.82 & 60.85 & 33.85 & 13.16 & 188.45 \\
          & Dadu & 161.11 & 40.74 & 180.86 & 10.33 & 93.25 & 41.13 & 75.34 & 133.20 \\
          & Larkana & 322.26 & 55.97 & 19.78 & 16.06 & 43.05 & 15.72 & 8.45 & 202.46 \\
          & Moro & 144.32 & 64.67 & 217.60 & 27.43 & 138.24 & 34.63 & 76.26 & 165.20 \\
         
         \hline
        \end{tabular}
    \label{fig:crop_damage2}
\end{table*}

To identify built-up areas impacted by floods, we used ESA's WorldCover v100 imagery with a 10-meter resolution. This imagery covered various features such as highways, city centers, and buildings on urban outskirts. By overlaying this data on the flood-affected imagery, we precisely determined the extent of damage to the built environment.

Additionally, we examined the potential influence of high soil moisture levels on Rabi crop cultivation. Comparing soil moisture data from January 2022 and January 2023, we assessed the impact on Rabi crop productivity. We used NASA's SMAP level 4 product to measure soil moisture levels in both the surface (0–5 cm) and root zone (0-100 cm) over an eight-month period from June 2022 to January 2023. This comprehensive approach enabled a detailed analysis of the flood's effects on crop cultivation and production in Sindh Province.

\section{Evaluation Results}
\subsection{Time Series Modelling for NDVI values}
\begin{figure}[!h]
    \centering
        \begin{tabular}{c} 
        \includegraphics[width=0.5\textwidth] 
            {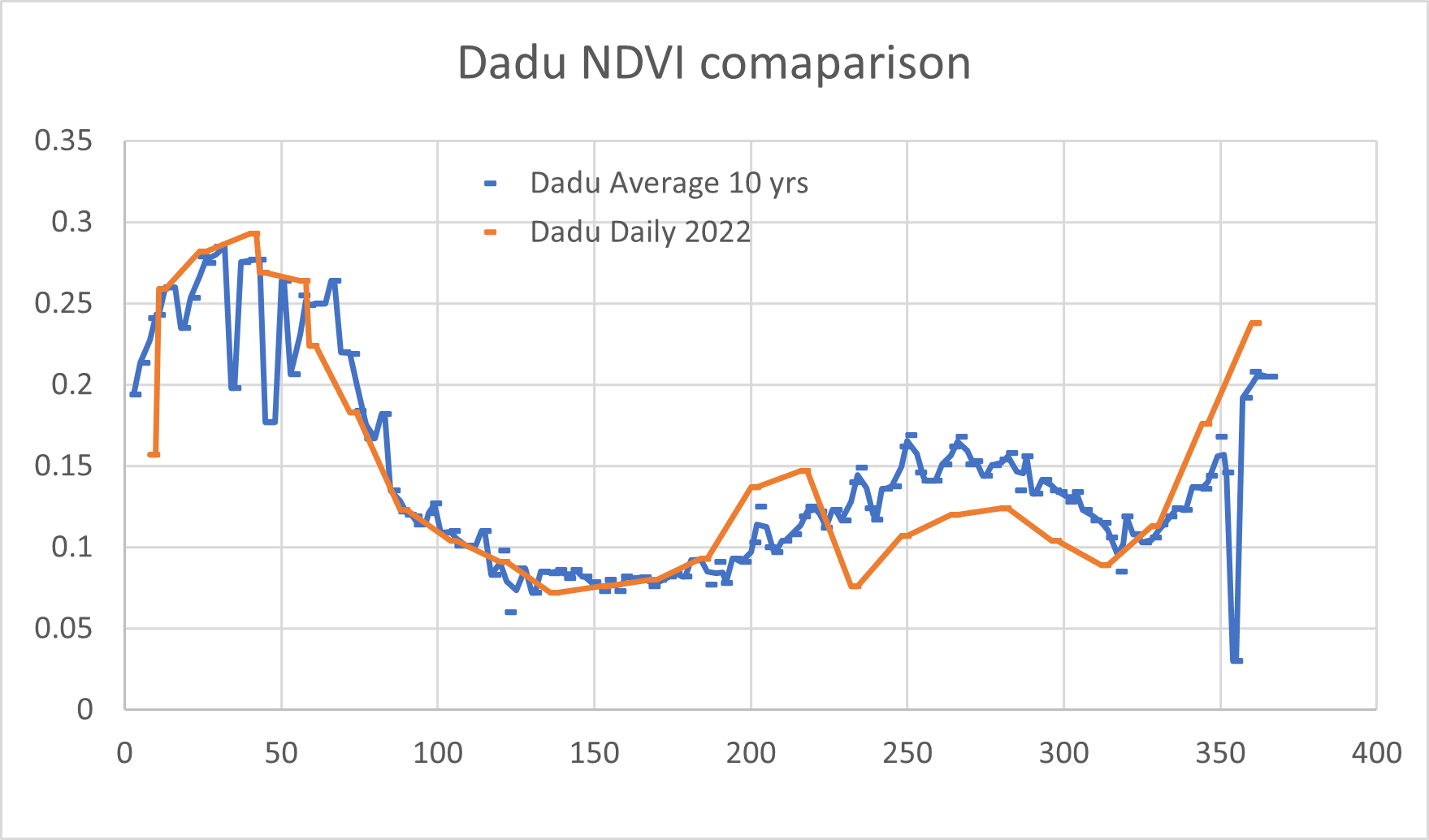} \\
        Dadu \\[5pt] 
        
        \includegraphics[width=0.5\textwidth]
            {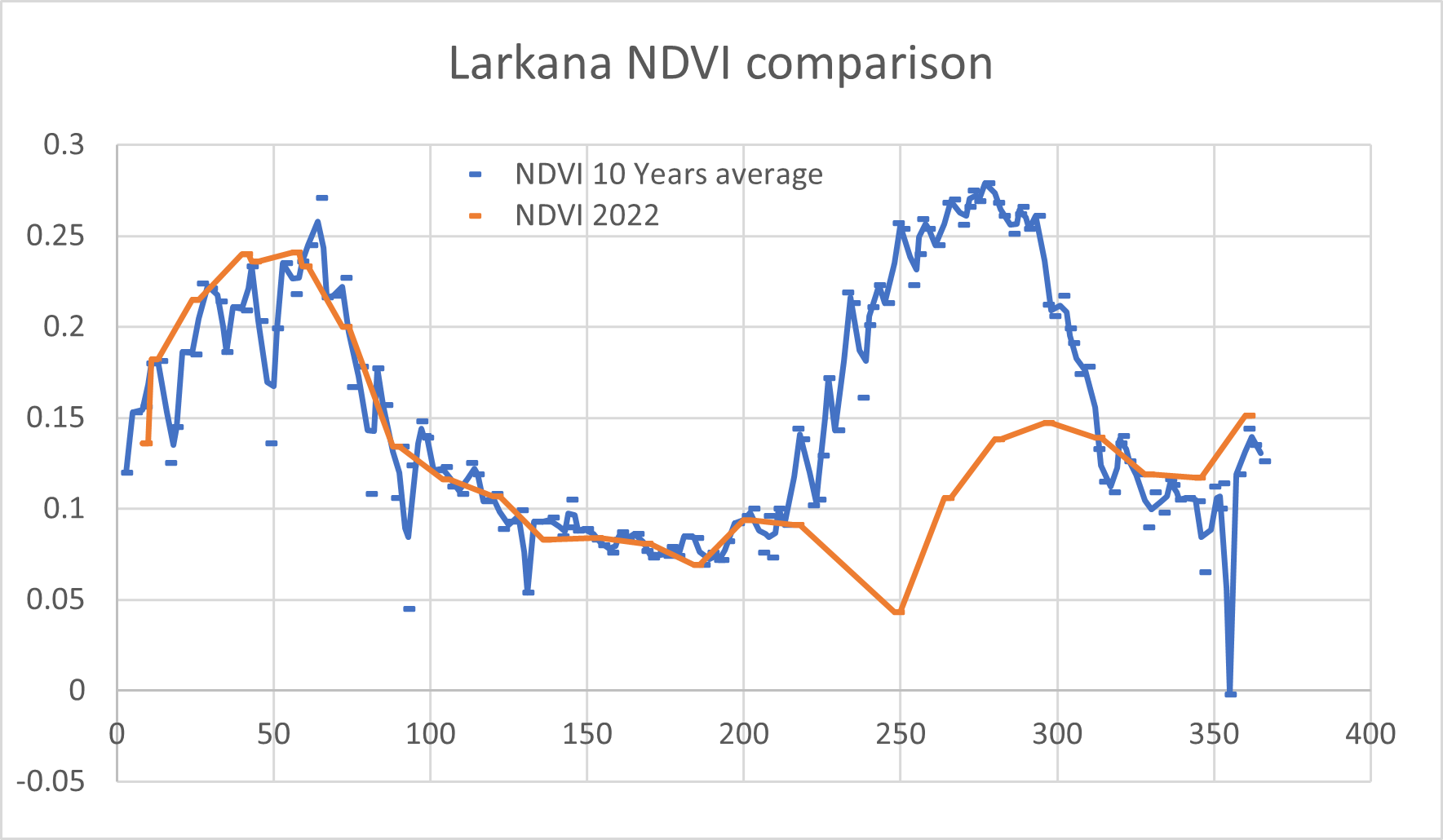} \\
        Larkana \\
        \end{tabular}
    \centering
    \caption{Time Series Modelling for NDVI values of the two tehsils of Sindh}
    \label{fig:ndviGraphs}
\end{figure}

Time series modeling for NDVI values of the two tehsils in Sindh can help identify and understand how vegetation in these regions evolves, especially in response to significant events like the floods that occurred in 2022 in Pakistan. By analyzing NDVI data over time and comparing it to the timing and severity of the floods, researchers can assess the impact of the flood event on vegetation health and recovery. This information can be crucial for disaster management, land-use planning, and environmental monitoring efforts to mitigate the effects of such natural disasters in the future (see Fig.~\ref{fig:ndviGraphs}).

\subsection{Quantitative Analysis}
Table~\ref{fig:crop_damage} displays the comprehensive data on affected Kharif crops and built-up areas across four districts within the study area. Among these districts, Mehar stood out as the most severely impacted, registering 129.9 sq km of affected crops and 0.70 sq km of impacted built-up regions. 
Table~\ref{fig:crop_damage2} presents data pertaining to the production areas of significant crops in each tehsil.

\begin{table*}[h]
\centering
\caption{Performance comparison of classification models.}
\begin{tabular}{|c|c|c|c|c|c|}
\hline
\textbf{Model Name} & \textbf{Training Accuracy} & \textbf{Validation Accuracy} & \textbf{Recall} & \textbf{F1 Score} & \textbf{Precision} \\ \hline
Random Forest \cite{breiman2001random} & 0.96 & 0.94 & 0.92 & 0.96 & 0.97 \\ \hline
SVM \cite{hearst1998support} & 0.72 & 0.71 & 0.45 & 0.56 & 0.57 \\ \hline
Naive Bayes \cite{dempster1977maximum} & 0.45 & 0.44 & 0.001 & 0.62 & 0.45 \\ \hline
Minimum Distance \cite{duda1973pattern} & 0.44 & 0.45 & 0.43 & 0.26 & 0.29 \\ \hline
\end{tabular}
\label{Models-performance-table}
\end{table*}



        
        
\begin{figure*}
    \centering
    \includegraphics[width=0.93\textwidth]{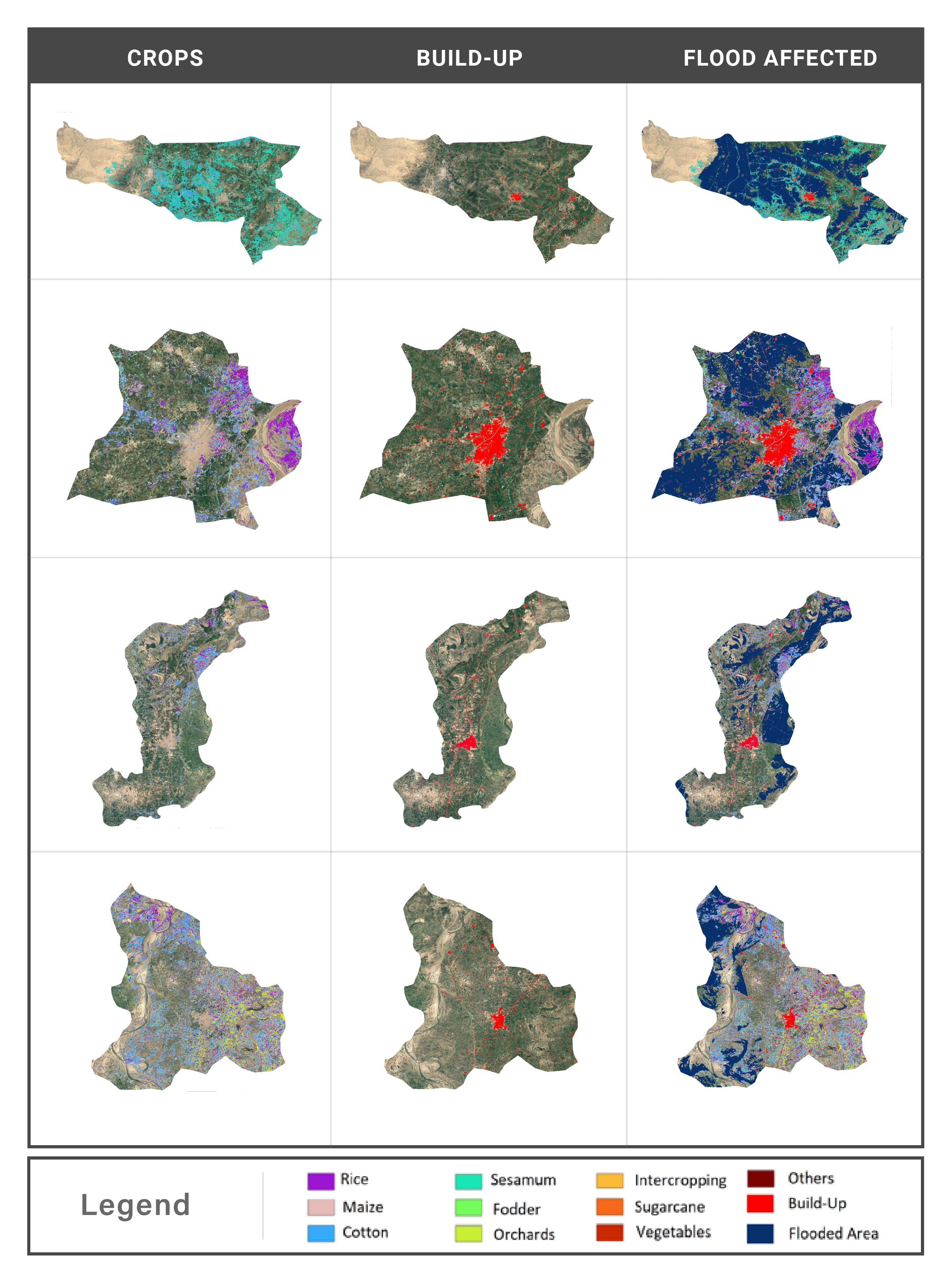}
   \caption{Crops and Built-up Area Affected By Floods in Pakistan 2022.}
        \label{fig:affected_cropsandBuilt-up}
\end{figure*}
To quantify the decrease in Rabi season productivity resulting from delayed cultivation, we conducted a comparative analysis between January 2022 and 2023. We assessed soil moisture levels separately for the root and top layers over an eight-month period in 2022, utilizing NASA's SMAP level 4 product. This involved gathering soil moisture maps for both the top layer (0–5 cm) and the root zone (0-100 cm) from June 2022 to January 2023, as illustrated in Fig.~\ref{fig:soil_moisture}. Similarly, we adopted a comparable methodology for Rabi crop classification, focusing on the period from October to November and April to May. To evaluate the decline in productivity in 2023 compared to 2022, we obtained the Rabi crop dataset from the Punjab Irrigation Department. The analysis incorporated soil moisture imagery to discern that the elevated soil moisture levels until January 2023 may have contributed to the delayed cultivation of Rabi crops. Figure~\ref{fig:affected_cropsandBuilt-up} visually represents the flood's impact on these regions.

The Table~\ref{Models-performance-table} provides a detailed comparison of crop classification results across various classifiers. It includes essential performance metrics for each model, such as the Model Name, which identifies the specific classifier used. The Training Accuracy and Validation Accuracy metrics reflect how well each model performs on the training and validation datasets, respectively, giving insights into their ability to generalize. The table presents Recall, which measures the model’s capacity to identify all relevant instances of the target class, and Precision, indicating the accuracy of positive predictions made by the model. The F1 Score is also included, offering a balanced evaluation by combining Precision and Recall into a single metric. This comprehensive analysis enables a thorough assessment of each classifier’s effectiveness and provides a clear view of their respective strengths and weaknesses in crop classification tasks.


\subsection{Qualitative Results}
The crop classification images within the four districts (Mehar, Dadu, Larkana, and Moro), as previously specified within the area of interest, are illustrated in Figure~\ref{fig:affected_cropsandBuilt-up}.

\section{Conclusion}
The study's key insights highlight the severe consequences in Pakistan's Sindh province, particularly the extensive loss of Kharif season crops typically planted between March and November. Additionally, post-flood high soil moisture content, as determined by SMAP satellite data, caused substantial delays in the cultivation of Rabi crops. These findings provide vital information for decision-makers and stakeholders involved in flood risk management and disaster response, offering essential guidance for future preparedness and resilience strategies. 

\section{Future Work}
Looking ahead, this research illuminates the detrimental consequences of floods on crop cultivation and production in the Sindh region of Pakistan. Leveraging Landsat 9 imagery and soil moisture maps, this study has yielded valuable insights into the extent of flood-induced damage and its enduring effects on crop production. The comparison of the NDVI data from the last decade with the post-flood production of 2022 reveals a dramatic decline in crop health, as depicted in Fig.~\ref{fig:ndviGraphs}. These discoveries serve as a clarion call to policymakers and stakeholders, underscoring the imperative of enhancing flood management strategies to mitigate the repercussions of such natural disasters on crop yields and food security.

Furthermore, the results demonstrate that the flood of 2022 profoundly impacted most Kharif crops and resulted in a substantial delay in the cultivation of Rabi crops. This delay was primarily attributed to the excessive moisture levels stemming from the floods, as substantiated by the soil moisture maps. Moreover, our findings suggest that due to these elevated moisture levels, the water requirements for crops during the ensuing Rabi season are anticipated to be lower compared to the previous year. This insight will guide future regional agricultural planning and water resource management efforts.

\bibliography{refs}

\end{document}